\documentclass[a4paper,12pt]{article}
\usepackage{exscale}
\usepackage{amssymb}
\usepackage{latexsym}

\newcommand{\ssection}[1]{\setcounter{equation}{0}\setcounter{footnote}{0}%
\section{#1}}

\newcommand{\equ}[2]{\begin{equation}\label{e#1}#2\end{equation}}
\newcommand{\equa}[2]{\begin{eqnarray}\label{e#1}#2\end{eqnarray}}
\newcommand{\D}{\displaystyle}
\newcommand{\mr}{\mathrm}
\newcommand{\gam}{\mathrm{\Gamma}}

\renewcommand{\(}{\left(}
\renewcommand{\)}{\right)}
\renewcommand{\[}{\left[}
\renewcommand{\]}{\right]}
\newcommand{\<}{\left\{}
\renewcommand{\>}{\right\}}
\renewcommand{\^}{\hat}
\renewcommand{\phi}{\varphi}
\renewcommand{\theta}{\vartheta}
\newcommand{\BBox}{{\rule{0cm}{1.5ex}\Box}}
\newcommand{\del}{\bigtriangleup}
\newcommand{\nab}{\nabla}
\newcommand{\iint}[1][z]{\int\!\! d^{\;\!8} #1\:}
\newcommand{\vint}[1][z]{ \int \;\!\!\! d^{\;\!8} #1 \:\mr{tr}\,}
\newcommand{\chint}[1][z]{\int \;\!\!\! d^{\;\!6} #1 \:\mr{tr}\,}
\newcommand{\lie}[1][\!\(c, \bar c\)]{{\cal L}_{\frac{1}{2}\,V} #1}
\newcommand{\Tr}{\mr{Tr}}
\newcommand{\JJ}{{\cal J}}
\newcommand{\eps}{\varepsilon}

\begin{document}
\title{ Effective Average Action in N=1~Super-Yang-Mills Theory}
\vspace{-1ex}
\author{\large{\bf{Sven~Falkenberg\thanks{
        e-mail: Sven.Falkenberg@itp.uni-leipzig.de}}~ 
        and \bf{Bodo~Geyer\thanks{
        e-mail: geyer@rz.uni-leipzig.de}}}\\
        \emph{Universit\"at~Leipzig, Institut~f\"ur~Theoretische~Physik}\\
        \emph{Augustusplatz~10-11, D-04109~Leipzig, Germany}}

\date{}
\maketitle
\begin{abstract}
\noindent
For N=1 Super-Yang-Mills theory we generalize the effective average action 
$\Gamma_k$ in a manifest supersymmetric way using the superspace formalism. 
The exact evolution equation for $\Gamma_k$ is derived and, introducing as 
an application a simple truncation, the standard one-loop $\beta$-function 
of N=1 SYM theory is obtained.
\end{abstract}


\ssection{Introduction}
Supersymmetric quantum field theories have attracted much attention in the 
last years. Some of them are low energy effective theories of superstring 
theories. In addition, the vacuum of the D=11 supergravity is part of the 
moduli space of consistent vacua of the so-called M-theory which can be 
considered as a more fundamental theory whose different representations are 
various superstring theories and a D=11 supergravity; they are related to 
each other by some duality transformations \cite{r}.

It is natural that one wants to have not only a perturbative approach to 
supersymmetric theories. A possible framework to study also nonperturbative 
properties of these theories is provided by Wilson's renormalization group 
\cite{b}. 
A special application of Wilson's renormalization group approach is given by 
the effective average action $\gam_k$ introduced by Wetterich and Reuter 
\cite{d}. In this approach the path integral is regularized by an UV- and an 
IR-cutoff. In the case when the theory is renormalizable one may send the 
UV-cutoff scale $\Lambda$ to infinity. The IR-cutoff scale $k$ is necessary 
to regularize the propagators of massless fields by a scale-dependent 
masslike term. This "mass" vanishes in the UV-region such that the theory 
remains unchanged there. By construction, in the limit $k \to 0$  the 
standard generating functional of the 1PI-Green's functions $\gam$ is 
obtained. On the other hand, in the limit $k \to \infty$ the effective 
average action becomes the classical action $S_\mr{cl}$. The evolution of 
$\gam$ with respect to the scale $k$ can be understood as increasing 
integrating-out of modes with momenta $p^2 > k^2$ (or averaging over these 
modes).

In this sense, $\gam_k$ interpolates between $S_\mr{cl}$ and $\gam$ 
describing an effective theory at scale $k$. The dependence from the scale 
$k$ is controlled by an exact evolution equation (the renormalization group 
or flow equation). Since it is a first order differential equation, from 
the knowledge of some initial value $\gam_\Lambda$ one can determine 
$\gam_k$ for all $k<\Lambda$. In general, this evolution equation cannot 
be solved exactly. However, approximate solutions can be obtained by the 
method of truncation. In the next sections we will explain this in detail. 
Despite the truncation being only an approximation one usually not only 
obtains the results of the standard perturbation theory but also 
nonperturbative contributions.

The method of effective average action has been applied successfully to 
different aspects of gauge theories \cite{d, e, f, g, n} and to 
gravity (being treated as a gauge theory) \cite{k}. Of course, it is 
desirable to generalize the method of effective average action also to 
the case of supersymmetric theories. A first attempt in this direction 
was the paper of Granda~et~al.~\cite{l} where a N=4 supergravity has been 
considered. The formalism used there is the same as has been used in 
\cite{k} for Einstein gravity. Consequentially, since supersymmetry is 
broken, the result differs from that which would be obtained from a manifest 
SUSY theory. 

Our goal in this paper is to maintain SUSY on every stage of the calculation. 
Therefore, it is convenient to use the superspace formalism \cite{a, p}. As 
an example we consider pure N=1 Super-Yang-Mills (SYM) theory in four 
dimensions. The case of usual, non-supersymmetric YM theories is treated in 
\cite{e} and extended in \cite{f, g}.

The paper is organized as follows. In the section~2 we define the effective 
average action $\gam_k$ in N=1 SYM theory. We work in the superspace and use 
the conventions of \cite{a}. Then, in section~3, we analyse the 
BRST-properties of $\gam_k$. The fourth section is devoted to the derivation 
of the exact evolution equation controlling the scale dependence of $\gam_k$. 
To find approximate solutions we restrict the space of admissible actions 
$\gam_k$ and obtain a truncated evolution equation. In section~5, as an 
application, we introduce a simple truncation and compute the one-loop 
$\beta$-function of N=1 SYM theory. The obtained result is in accordance with 
the well-know expressions found in the literature \cite{o}.


\ssection{Effective Average Action}
In order to derive the effective average action in N=1 SYM let us consider
the following generating functional\footnote{To obtain the Euclidean version
of the Minkowskian action according to \cite{s} a continuous Wick rotation 
is performed where, however, also the spinor structure simultaneously has to 
be rotated. For Weyl spinors one has to drop the hermiticity requirement 
which is in Minkowski space 
$\psi^\dagger_\mr{L} = \psi^\dagger \, \frac{1 + \gamma_5}{2}$ if
$\psi_\mr{L} = \frac{1 + \gamma_5}{2} \, \psi$. The path integral is not 
affected by this since $\psi$ and $\psi^\dagger$ are always treated as 
independent fields. As required one obtains the analytic continuations of the 
Green's functions.}
\equ{2.1}{Z_k [j_V, J, \JJ; \Omega] \equiv e^{W_k [j_V, J, \JJ; \Omega]} 
          = \int \!\! {\cal D} V {\cal D} \Phi \,
            e^{- S_k[V, \Phi, j_V, J, \JJ; \Omega]} }
where  $S_k$ is some action being invariant under, e.\,g., the gauge group 
$SU(N)$; it depends on the gauge field $V$ having the source  $j_V$, on the 
(anti-) ghost fields $\Phi$ with sources $J$, on the BRST-sources $\JJ$ as 
well as on the background gauge field $\Omega$.

Here we use the background field method \cite{a, c} where the complete gauge 
field ${V_\mr{S}}$ is splitted into two background parts $\Omega$ and 
$\bar \Omega$ and the fluctuations $V$ around this background:
\equ{2.2}{e^{V_\mr{S}} = e^\Omega e^V e^{\bar\Omega}.}
For non-abelian gauge groups this splitting is highly nonlinear. It allows
to write the gauge transformation of the complete gauge field in different
manners. Let us first introduce covariant SUSY derivatives in a background
vector but quantum chiral representation \cite{a, p},
\newline
\parbox{13.9cm}{$$\begin{array}{r@{\:=\:}l@{,\quad}r@{\:=\:}l@{\quad\;}l}
         \nab_\alpha               & e^{-\Omega} D_\alpha e^\Omega &
         \bar\nab_{\dot\alpha}     & e^{\bar\Omega} \bar D_{\dot\alpha}
         e^{-\bar\Omega}           &
            \mbox{background covariant derivatives,}                  \\
         {\cal D}_\alpha           & e^{- V} \nab_\alpha e^V       &
         \bar{\cal D}_{\dot\alpha} & \bar \nab_{\dot\alpha}        &
            \mbox{total covariant derivatives}
                  \end{array}$$}\hfill
\parbox{7mm}{\equa{2.3}{}}
\newline
where $D_\alpha$ and $\bar D_{\dot \alpha}$ are the usual SUSY derivatives.
The gauge transformation of $V_\mr{S}$ reads
$$\(e^{V_\mr{S}}\)' = e^{i \bar \Lambda_0} e^{V_\mr{S}} e^{-i \Lambda_0}$$
with a chiral superfield $\Lambda_0, \bar D^{\dot \alpha} \Lambda_0 = 0$.
It can be written as a transformation acting on $V$ only but leaving $\Omega$ 
(and $\bar\Omega$) invariant (\emph{quantum transformation}):
\equ{2.4}{\(e^{V_\mr{S}}\)' =  e^\Omega \(e^{-\Omega} e^{ i \bar \Lambda_0} 
            e^\Omega   e^V     e^{\bar \Omega  }      e^{-i      \Lambda_0} 
            e^{-\bar \Omega}\) e^{\bar \Omega}.}
In this case the covariant derivatives transform as 
($A = \{\alpha, \dot\alpha, \alpha \dot\alpha\}$)
$$ \nab_A'     = \nab_A,\quad
   {\cal D}_A' = e^{i \Lambda_0} {\cal D}_A\, e^{-i \Lambda_0}.$$
As a second possibility we require the fluctuations $V$ to transform
homogeneously (\emph{background transformation}):
\equ{2.5}{\(e^{V_\mr{S}}\)' = \(e^{i \bar \Lambda_0} e^\Omega e^{-i K}\)
          \Big(e^{i K} e^V e^{-i K}\Big)\( e^{i K} e^{\bar \Omega} e^{-i
          \Lambda_0}\).}
Here $K$ denotes a real superfield which is related to $\Lambda_0$ in the 
usual way (see \cite{a, c}). We conclude that, indeed, 
$V' = e^{i K} V e^{-i K}$ transforms homogeneously. Then the derivatives have 
the following transformation laws
$$          \nab_A'   = e^{i K} \nab_A e^{-i K}                   ,\quad
          {\cal D}_A' = e^{i K} {\cal D}_A \,e^{-i K}. $$
As a third possibility one can require that the background transforms
homogeneously; however, we will not consider this case.

Now we specify the various parts of the action
$$ S_k = S + S_\mr{gf} + S_\mr{gh} + \del_k S + S_\mr{sou}.$$
The classical gauge invariant action $S$ reads
\equ{2.6}{S [V; \Omega] =  \frac{1}{g^2} \chint W^\alpha W_\alpha
                        = -\frac{1}{2 g^2} \vint \!\!\[ \(e^{-V} \nab^\alpha
                        e^V\) \cdot \bar \nab^2\!\! \(e^{-V} \nab_{\!\alpha}
                        e^V\) \] }
with the coupling constant $g$. The integral over the chiral space has been 
denoted by $\int \!d^{\;\!6} z \equiv \int \!d^{\;\!4} x \, d^{\;\!2} \theta$, 
while $\int \!d^{\;\!8} z \equiv \int \!d^{\;\!4} x \, d^{\;\!2} \theta \, 
d^{\;\!2} \bar\theta$; the trace $\mr{tr}$ has to be taken with respect to the 
gauge group. 

As gauge fixing term for the fluctuation $V$ we choose
\equ{2.7}{S_\mr{gf} [V; \Omega] = -\frac{1}{2 \alpha g^2} \vint V 
          \(  \nab^2 \bar \nab ^2 + \bar \nab^2 \nab^2\) V}
with gauge parameter $\alpha$. This term is still invariant under the 
background transformation (\ref{e2.5}) because of the homogeneity of the 
transformation for $V$. 

To this gauge fixing belongs the following ghost term
\newline
\parbox{13.9cm}{\begin{eqnarray*}
                S_\mr{gh} [V, C; \Omega] 
          &=& - \vint \!\!\< \(c' + \bar c'\) 
                \lie[ \[\(c + \bar c\) + \coth \lie[]\(c-\bar c\)\] ] + 
                \bar b b\>                                          \\
     &\equiv& - \vint \!\!\< \(c' + \bar c'\) 
                \lie + \bar b b\>  
                  \end{eqnarray*}}\hfill
\parbox{7mm}{\equa{2.8}{}}
\newline
with the (anti-) ghost fields 
$C \equiv \{c, c', \bar c, \bar c', b, \bar b\}$. The fields $c$, $c'$ and 
$b$ are background chiral whereas $\bar c$, $\bar c'$ and $\bar b$ are 
background anti-chiral fields; by ${\cal L}$ we denote the Lie-derivative 
${\cal L}_V [A] \equiv \[V, A\]$.

It is convenient to solve the chirality constraint and to introduce the 
\mbox{(anti-)} ghost prepotentials 
$\Phi \equiv \{\psi, \psi', \bar\psi, \bar\psi', \phi, \bar\phi\}$ according 
to
$$ \{c, c', b\} = \bar \nab^2 \{\psi, \psi', \phi\}, \qquad
   \{\bar c, \bar c', \bar b\} = \nab^2 \{\bar \psi, \bar \psi', 
   \bar \phi\}.$$
Thereby we introduced an additional gauge freedom for the (anti-) ghosts
\equ{2.8.1}{\delta \psi = \bar \nabla^{\dot\alpha} \bar
            \omega_{\dot\alpha} \quad\ldots}
with some arbitrary superfield $\bar \omega_{\dot\alpha}$. We do not gauge 
this symmetry here.

The term which is essential for the effective average action is the cutoff
term $\del_k S$. It is a masslike damping term which provides the
IR~regularization of the theory:
\newline
\parbox{13.9cm}{$$\begin{array}{r@{}l}
       \D   \del_k S [V, \Phi; \Omega] = \vint \bigg[
     & \D             V    R_k^\mr{YM}\!\(-\frac{\BBox_V}{k^2}\) V
         +      \bar\psi'  R_k\!        \(-\frac{\BBox_+}{k^2}\) \psi \\[2.2ex]
     & \D+  \:  \bar\psi   R_k\!        \(-\frac{\BBox_+}{k^2}\) \psi'\!
         +      \bar\phi\, R_k\!        \(-\frac{\BBox_+}{k^2}\) \phi \bigg]
                  \end{array}$$}\hfill
\parbox{7mm}{\equa{2.9}{}}
\newline
where $\BBox_V$ and $\BBox_+$ are background covariant generalized Laplacians.
$R_k$ and $ R_k^\mr{YM}$ are cutoff operators with the cutoff scale $k$ 
\cite{d}. They are proportional to a dimensionless damping function
$R_k \sim k^2 R^{\(0\)}$ which smoothly interpolates between $0$ and $1$:
$$ \lim_{k^2\to 0} R^{(0)} = 0 , \qquad  \lim_{k^2 \to \infty} R^{(0)} = 1
   \qquad \mbox{(p fixed)}.$$
A convenient choice for the damping function is
\equ{2.9.1}{R^{(0)}\!\(p^2/k^2\) = \frac{p^2/k^2}{e^{\,p^2/k^2} - 1}.}
Since the cutoff term $\del_k S$ is bilinear it modifies the propagators of 
the theory in such a way that the free propagator is replaced by
$$ \frac{1}{p^2} \to \frac{1}{p^2 + R_k \!\(p^2/k^2\)}.$$
This cutoff acts as a scale dependent mass vanishing for $p^2/k^2 \to \infty$. 
In the IR region $p^2 < k^2$ the cutoff is a mass playing the role of a 
regularizator. In principle one could choose different cutoffs for the 
different types of ghost fields but it turns out that there is no reason to 
do that. Therefore we use the same $R_k$ for all ghosts. Of course, by 
construction, from $R_{k \to 0} = 0$ it follows that $\del_{k \to 0} S = 0$. 

In addition we introduce the source term $S_\mr{sou}$ for the fields
\equ{2.10}{S_\mr{sou}[V, \Phi, j_V, J, \JJ;\Omega] = - \vint \!\! \< j_v V + 
           J \Phi +  \tau \lie + \sigma \psi \bar\nab^2 \psi +
             \bar\sigma \bar\psi \nab^2 \bar\psi\>}
where $J = \{j_\psi, \bar j_{\bar\psi}, j_{\psi'}, \bar j_{\bar\psi'},
j_\phi, \bar j_{\bar\phi}\}$ are the sources to the ghost prepotentials $\Phi$.
Here we also introduced the BRST-sources $\JJ = \{\tau, \sigma, \bar\sigma\}$
which couple to the nonlinear BRST-transforms of $V, \psi$ and $\bar\psi$,
respectively. 

For a proper definition of (\ref{e2.1}) one needs an UV-regularization. Of 
course, one would prefer a regularization which respects the symmetries of 
the theory, like BPHZ (see \cite{q}). Then one may send the UV-cutoff 
$\Lambda \to \infty$. However, if such a scheme does not exist one chooses 
$\Lambda$ much larger than every other physical scale of the theory but 
finite. Typically,  in that case all of the relations are valid up to 
$O (1/\Lambda)$.

After defining the generating functional we perform a Legendre transformation.
Let us first introduce the \emph{k-dependent} mean fields
\equ{2.11}{v   := \frac{\delta W_k}{\delta j_V}, \quad
          \Xi  := \frac{\delta W_k}{\delta J}}
where $\Xi = \{\xi, \xi', \bar\xi, \bar\xi', \beta, \bar\beta\}$. Then, the
effective average action is defined as \cite{d}
\equ{2.12}{\gam_k[v, \Xi, \JJ; \Omega] := \vint \!\[ j_V v + J \Xi \] -
              W_k[j_V, J, \JJ; \Omega] - \del_k S [v, \Xi; \Omega] .}
As is well known the cutoff must be subtracted from the usual Legendre 
transform because $\gam_k$ should fulfill the boundary conditions
$\gam_{k \to \infty} = S_\mr{cl}$ and $\gam_{k \to 0} = \gam$.
Then this effective average action describes an effective theory at scale $k$
averaging over modes with $p^2 > k^2$. It smoothly interpolates between the
classical and the usual effective action.

Because the cutoff vanishes for $k \to 0$ it is evident that 
$\gam_{k \to 0} = \gam$. Let us now analyse the limit $k \to \infty$.
From the definition of $\gam_k$ it follows that
$$e^{-\gam_k} = \exp\< -\vint \[j_V v + J \Xi\] + W_k + 
  \del_k S[v, \Xi; \Omega]\>$$
which  after some further steps leads to the following integral equation
$$
\begin{array}{r@{\,}c@{\,}l}
\D e^{-\gam_k} = \int \!\! {\cal D} V {\cal D} \Phi \, \exp \!\bigg\{
          &-& \D S_\mr{cl}\, + \,\vint \!\[\(V-v\)
              \frac{\delta \gam_k}{\delta v}  +
              \(\Phi - \Xi\)\frac{\delta \gam_k}{\delta \Xi}\]   \\
          &-& \D \del_k S\[V-v, \Phi - \Xi; \Omega\]\bigg\}
\end{array}
$$
with $S_\mr{cl} = S + S_\mr{gf} + S_\mr{gh} + S_\mr{sou,BRST}$. The exponential 
$\exp(- \del_k S)$ containing the cutoff operators $R_k$ becomes proportional 
to a (generalized) Dirac $\delta$-functional $\delta[V-v] \,\delta[\Phi-\Xi]$ 
equating quantum and mean fields\footnote{This analysis can be done in a more 
accurate way using the saddle point method.}. Therefore, we find the required 
result
\equ{2.13}{\gam_{k \to \infty} = S_\mr{cl}}
ensuring that $\gam_k$ is interpolating between $S_\mr{cl}$ and $\gam$.
                                 

\ssection{BRST-Properties of $\gam_k$}
One of the drawbacks of the effective average action is that the cutoff terms 
break BRST-invariance. This is also the case for the supersymmetric 
extension. It leads to a modification of the Slavnov-Taylor-identity (STI) 
\cite{f, i} which now will be specified.

To obtain a nilpotent BRST-transformation we modify the gauge fixing to 
\cite{q} 
\equ{3.1}{S_\mr{gf}^{\(\alpha, B\)} [V, B; \Omega] = -\frac{1}{g^2} \vint\!\! 
          \( -\alpha B \bar B + B \nab^2 V + \bar B \bar \nab^2 V\)}
by introducing Lagrangian multiplier fields $B$ and $\bar B$. Their Legendre 
transform will be denoted by $\tilde b$ and $\bar{\tilde b}$. In addition we 
need sources for $B$ and $\bar B$:
\equ{3.2}{S_\mr{sou}^{\(\alpha,B\)} = - \vint\!\! \< j_v V + J \Phi + j_B B +
             \bar j_{\bar B} \bar B + \tau \lie +
             \sigma \psi \bar\nab^2 \psi +
             \bar\sigma \bar\psi \nab^2 \bar\psi\>.}
Expressing all terms with the help of prepotentials the sum
$S + S_\mr{gf}^{\(\alpha, B\)} + S_\mr{gh}$ is invariant under the following
nilpotent BRST-transformation
\newline
\parbox{13.9cm}{$$\begin{array}{r@{\:=\:}l@{,\qquad}r@{\:=\:}l}
          \D  s  V        & \D  g\, \lie                                \\
          \D  s \psi      & \D -g\, \psi \bar \nab^2     \psi     &
          \D  s \bar\psi  & \D -g\, \bar\psi  \nab^2     \bar\psi,      \\
          \D  s \psi'     & \D -\frac{1}{g}\,            \bar B   &
          \D  s \bar\psi' & \D -\frac{1}{g}\,                 B,        \\
          \D  s  B        & \D 0                                  &
          \D  s \bar B    & \D 0,                                       \\
          \D  s \phi      & \D 0                                  &
          \D  s \bar\phi  & \D 0   .
                  \end{array}$$}\hfill
\parbox{7mm}{\equa{3.3}{}}
According to the transformations of $S_\mr{sou}$ and $\del_k S$ the STI 
obtains a symmetry breaking term $\del_k$ being proportional to the cutoff 
operators. In terms of $\gam_k' = \gam_k - S_\mr{gf} $ the modified STI 
reads
\equ{3.4}{\vint\!\[\frac{\delta \gam_k'}{\delta       v} \frac{\delta \gam_k'}{
\delta  \tau} -    \frac{\delta \gam_k'}{\delta     \xi} \frac{\delta \gam_k'}{
\delta\sigma} -    \frac{\delta \gam_k'}{\delta \bar\xi} \frac{\delta \gam_k'}{
\delta\bar\sigma} \] = \del_k .}
To derive it we used the anti-ghost equations of motion
$$\bar\nab^2 \! < \lie > + R_k \bar\xi = j_{    \psi'},$$
$$    \nab^2 \! < \lie > - R_k     \xi = j_{\bar\psi'} $$
where $<\!A\!> \, \equiv Z_k^{-1} \int{\cal D} V {\cal D} \Phi \,A\, e^{-S_k}$.
The term $\del_k$ can be written in the following form
\newline
\parbox{13.9cm}{\begin{eqnarray*}
           \del_k  =
 &-&  2\;\:\;
      \Tr' \[\(\gam_k^{[2]} + \^R_k\)_{\!v       \^\Xi }^{-1} \frac{\delta^2
           \gam_k}{\delta \tau \delta   \^\Xi} \,\(\^R_k\)_{\!    v      v}\] \\
 &-&  \quad\;
      \Tr' \[\(\gam_k^{[2]} + \^R_k\)_{\!\bar\xi' \^\Xi}^{-1} \frac{\delta^2
           \gam_k}{\delta \sigma \delta \^\Xi} \,\(\^R_k\)_{\!\xi \bar\xi'}\] \\
 &+&  \quad\;
      \Tr' \[\(\gam_k^{[2]} + \^R_k\)_{\!\xi'     \^\Xi}^{-1} \frac{\delta^2
           \gam_k}{\delta \bar\sigma \delta \^\Xi}
                                             \,\(\^R_k\)_{\!\xi' \bar\xi}\]   \\
 &-&  \frac{1}{g^2}\,
      \Tr' \[\(\gam_k^{[2]} + \^R_k\)_{\!\bar\xi\,\^\Xi}^{-1} \frac{\delta^2
           \gam_k}{\delta \bar j_{\bar B}\,\delta \^\Xi}
                                             \,\(\^R_k\)_{\!\xi' \bar\xi}\]   \\
 &+&  \frac{1}{g^2}\,
      \Tr' \[\(\gam_k^{[2]} + \^R_k\)_{\!\xi  \,  \^\Xi}^{-1} \frac{\delta^2
           \gam_k}{\delta j_B   \,         \delta \^\Xi}
                                             \,\(\^R_k\)_{\!\xi \bar\xi'}\]. 
                \end{eqnarray*}}\hfill
\parbox{7mm}{\equa{3.5}{}}
\newline
Here we introduced a matrix notation which is very useful. 
$\(\gam_k^{[2]}\)_{\! A B}$ denotes the second functional derivative of 
$\gam_k$ with respect to the fields $A$ and $B$. With $\^\Xi$ we denote the 
set of \emph{all} mean fields $\^\Xi \equiv \{v, \Xi, \tilde b, 
\bar{\tilde b}\}$. 
Also the cutoff is written as a matrix in the space of fields:
$$\(\^R_k\)_{\!v v} \equiv R_k^\mr{YM}, \qquad \(\^R_k\)_{\!\xi' \bar\xi} =
  \(\^R_k\)_{\!\xi \bar\xi'} = \(\^R_k\)_{\!\beta \bar\beta}  \equiv R_k .$$
The trace in (\ref{e3.5}) has to be taken as $\Tr' \equiv \int\!\! d^{\;\!8}y 
\vint$.

For the functional $W_k$, eq.~(\ref{e2.1}) the breaking term $\del_k$ can be 
rewritten in a much simpler way:
\newline
\parbox{13.9cm}{\begin{eqnarray*}
           \del_k  &=&
  -   2\,\Tr \!\[   \frac{\delta^2 W_k}{\delta \tau       \,\delta j_V}  
  \,R^\mr{YM}_k\]  
  -      \Tr \!\[\( \frac{\delta^2 W_k}{\delta \sigma     \,\delta \bar 
  j_{\bar\psi'}}
  -                 \frac{\delta^2 W_k}{\delta \bar\sigma \,\delta j_{\psi'}} 
  \) R_k\]   \\
&&-  \frac{1}{g^2}\,
         \Tr \!\[\( \frac{\delta^2 W_k}{\delta \bar j_{\bar B}\,\delta \bar 
         j_{\bar\psi}}
  -                 \frac{\delta^2 W_k}{ \delta j_B           \,\delta j_\psi} 
  \) R_k\]. 
                \end{eqnarray*}}\hfill
\parbox{7mm}{\equa{3.6}{}}
Here the traces have to be taken over the gauge group and the superspace.
Because the cutoff operators decay exponentially for $p \to \infty$ the traces 
are UV-finite. In addition, $R_k$ and $R_k^\mr{YM}$ vanish for $k \to 0$ such 
that also $\del_k$ vanishes in this limit. Then (\ref{e3.4}) becomes the 
unbroken standard STI of N=1 SYM theory \cite{q}. Also in the UV region --- 
where the eigenvalues of $\BBox_V$ and $\BBox_+$ are much greater than the 
cutoff scale $k^2$ --- the breaking term $\del_k$ can be neglected. 

In \cite{q} the authors used a version of the BPHZ scheme to regularize the
theory. Since the gauge multiplet $V$ is dimensionless the scheme fails 
to remove all divergences: Some IR-divergences have to be removed by 
explicit introduction of a mass term for particular components of $V$. This 
has been done in a BRST-invariant but SUSY breaking way. In our approach 
such divergences are removed by the cutoff term. To see this let us consider
the $1/p^4$-part of a propagator being the worst situation causing the 
IR-singularities in \cite{q}. If such a term contributes to $\del_k$ in 
(\ref{e3.6}) it gives rise to an integral of the form
$$I_k \sim \int_0^\infty\!\!\mr{d} p \, p^2 \frac{1}{p^4}\, R_k (p^2);$$
the factor $p^2$ comes from the volume element. Using $R_k \sim k^2 R^{(0)}$ 
and (\ref{e2.9.1}) we obtain
$$ I_k \sim k \, \Gamma\!\(1/2\) \, \zeta\!\(1/2\)$$
where $\Gamma$ and $\zeta$ are the Gamma and Zeta function respectively. Thus
$I_{k \to 0} = 0$ as is required for the vanishing of $\del_k$ for $k \to 0$.

All that should not hide the fact that $\gam_k$ leads to a broken STI. 
Also the invariance of $\gam_k$ under the background transformation
(\ref{e2.5}), where the other fields transform homogeneously, cannot 
remedy this fact. To deal with this situation there are two possible 
strategies. First, one does not matter about the symmetry breaking for 
$k^2>0$, and one has to ensure that the unbroken identities are to be 
fulfilled for $k^2 = 0$ \cite{f, i}. For the non-supersymmetric effective 
average action this is considered in detail in \cite{f}. It is shown there 
that if $\gam_k$ fulfills (\ref{e3.4}) for some value $k=k_0$ it satisfies 
the modified STI for all $k=k'$ provided that $\gam_{k'}$ is obtained from 
$\gam_{k_0}$ by integration of the evolution equation which will be derived 
in the next section (see eq. (\ref{e4.1})). With this in mind the modified 
STI (\ref{e3.4}) can be used to control the breaking $\del_k$. 

On the other hand, it would be very desirable to have a manifest 
BRST-invariant formalism for any $k$. However, to our knowledge there exists 
no satisfactory consideration of that problem.


\ssection{Evolution Equations and Truncation}
Let us now write down the evolution equation for $\gam_k$. To this purpose we 
differentiate (\ref{e2.1}) with respect to the dimensionless parameter 
$t = \ln(k/k_0)$. Performing the Legendre transformation (\ref{e2.11}, 
\ref{e2.12}) we obtain the \emph{exact} evolution equation
\newline
\parbox{13.9cm}{\begin{eqnarray*}
          \partial_t \gam_k  
 &=&  \Tr \[\(\gam_k^{[2]} + \^R_k\)_{\!  v           v}^{-1}
                \(\partial_t \^R_k\)_{\!  v           v} \,\]
  -   \Tr \[\(\gam_k^{[2]} + \^R_k\)_{\!\xi   \bar\xi' }^{-1}
                \(\partial_t \^R_k\)_{\!\xi   \bar\xi' }\]   \\
 &-&  \Tr \[\(\gam_k^{[2]} + \^R_k\)_{\!\xi'  \bar\xi  }^{-1}
                \(\partial_t \^R_k\)_{\!\xi'  \bar\xi  }\]
  -   \Tr \[\(\gam_k^{[2]} + \^R_k\)_{\!\beta \bar\beta}^{-1}
                \(\partial_t \^R_k\)_{\!\beta \bar\beta}\] .
                \end{eqnarray*}}\hfill
\parbox{7mm}{\equa{4.1}{}}
\newline
The trace Tr has to be understood as integration over the superspace and as
usual trace over the gauge group. Equation (\ref{e4.1}) looks like an one-loop
equation for the effective action. But there are two important differences.
First, the traces are IR-regularized by the cutoff which appears as a scale
dependent mass here. In addition, the factor $\partial_t \^R_k$ regularizes
the traces also in the UV-region. Therefore the main contributions to the 
traces come from the modes with momenta around the scale $k$; no further 
UV-regularization is necessary. The second difference is the appearance of 
the \emph{full} propagator instead of the free one in the one-loop case. 
This fact shows the renormalization group improvement of the effective 
average action.

Equation (\ref{e4.1}) is a first order differential equation. Hence, if we 
know the initial value $\gam_\Lambda$ we can, in principle, integrate it to 
obtain $\gam_k$ for all $k < \Lambda$. Obviously, the information which is 
contained in  the evolution equation and in the STI is equivalent to the one 
in the generating functional (\ref{e2.1}). Consequently, from now on we 
define the theory with the help of (\ref{e4.1}) and (\ref{e3.4}) instead of 
(\ref{e2.1}). The functional (\ref{e2.1}) together with the symmetry 
properties were necessary to motivate the form of (\ref{e4.1}) and 
(\ref{e3.4}). They provide a simple physical interpretation of the  formalism.

Now the problem appears how to solve equation (\ref{e4.1}). In general this 
will be impossible or at least very difficult. However, truncating the 
space of possible action functionals provides a possibility to find at 
least \emph{approximate} solutions \cite{d}: One can consider $\gam_k$ as a 
sum of infinitely many terms containing an infinite number of generalized 
$k$-dependent coupling constants. These coupling constants span the 
parameter space of possible action functionals. $\gam_k$ is a trajectory in 
this space. A truncation means that one divides the set of coupling constants 
into two classes --- relevant and irrelevant ones with respect to the 
evolution. Then one makes an ansatz which contains only the few relevant 
ones neglecting the irrelevant ones.

We want to demonstrate this concept with the help of the following simple 
truncation:
\newline
\parbox{13.9cm}{\begin{eqnarray*}
\gam_k\!\[v, \Xi, \JJ; \Omega\]
           &=& \bar\gam_k\!\[\Omega = v_\mr{S}/2\]
            +    \^\gam_k\!\[v;      \Omega\]
            +  S_\mr{gf} \!\[v;      \Omega\]           \\
           &+& S_\mr{gh} \!\[v, \Xi; \Omega\]
            +  S_\mr{sou, BRST}\!\[v, \Xi, \JJ; \Omega\].
                \end{eqnarray*}}\hfill
\parbox{7mm}{\equa{4.2}{}}
\newline
The first term  $\bar\gam_k\!\[\Omega = v_\mr{S}/2\] \equiv
\gam_k\!\[v=0, \Xi=0, \JJ=0; \Omega=v_\mr{S}/2\]$ is that part of the 
effective action which contains the conrtibution of
$e^{v_\mr{S}} = e^\Omega e^{\bar\Omega}$. With $v_\mr{S}$ we introduced
the complete gauge mean field in analogy to $V_\mr{S}$ in (\ref{e2.2}):
\equ{4.3}{e^{v_\mr{S}} = e^\Omega e^v e^{\bar\Omega}.}

The second term $\^\gam_k\!\[v; \Omega\] \equiv
\gam_k\!\[v, 0, 0; \Omega \neq v_\mr{S}/2\]$ contains the contribution 
of $e^{v_\mr{S}} \neq e^\Omega e^{\bar\Omega}$. It vanishes for 
$e^{v_\mr{S}} = e^\Omega e^{\bar\Omega}$ because all contributions 
which do not vanish in this case are included in $\bar\gam_k$. This leads 
to the interpretation of $\^\gam_k$ as quantum correction to the gauge 
fixing term $S_\mr{gf}$ which vanishes also for 
$e^{v_\mr{S}} = e^\Omega e^{\bar\Omega}$. For the ghost term we have not 
included any quantum correction, i.\,e. we restricted ourselves to the 
tree approximation with respect to the ghosts.

If we look at the initial condition (\ref{e2.13}) we find that it is
fulfilled if
\equ{4.4}{\bar\gam_{k \to \infty} = S_\mr{cl} \qquad \mbox{and} \qquad
            \^\gam_{k \to \infty} = 0.}
Let us analyse the modified STI (\ref{e3.4}). It is evident that
$S_\mr{gf} + S_\mr{gh} + S_\mr{sou, BRST}$ is invariant under
BRST-transformations. Also $\bar\gam_k$ gives no contribution because it is 
independent of the sources, the ghosts and the field $v$. In the ansatz 
(\ref{e4.2}) only $\^\gam_k$ contributes to the STI which now has the 
following form:
\equ{4.5}{-\vint \frac{\delta \^\gam_k}{\delta v}\, \lie = \del_k.}

These properties of $\^\gam_k$ imply a further simplification: Let us neglect 
$\^\gam_k$ for all $k$. Thus the ansatz satisfies the unmodified STI with 
$\del_k = 0$ in (\ref{e3.4}). Neglecting the quantum corrections to the gauge 
fixing means neglection of $\del_k$. On the other hand, $\del_k$ has the form 
of loop integrals which give some corrections which presumably go beyond the 
truncation we use here. 

Summarizing the last considerations we ended up with the ansatz
\equ{4.5.1}{\gam_k = \bar\gam_k +  S_\mr{gf} +  S_\mr{gh} +  S_\mr{sou, BRST}}
providing a simple truncation to the effective average action. It is important 
that it fulfills the initial condition (\ref{e2.13}). Furthermore, it is 
distinguished by fulfilling the \emph{unmodified} STI. Therefore the STI 
implies no further restriction on $\gam_k$ in (\ref{e4.5.1}).

If we plug in the ansatz (\ref{e4.5.1}) into the exact evolution equation
(\ref{e4.1}) we obtain the following truncated evolution equation
\newline
\parbox{13.9cm}{\begin{eqnarray*}
          \partial_t \gam_k\!\[v; \Omega\]
 &=& \Tr\[\(   \gam_k^{[2]}\[v; \Omega     \] + \^R_k\)_{\!  v           v}^{-1}
          \(\partial_t \^R_k\)_{\!  v           v} \,\]                     \\
 &-& \Tr\[\(S_\mr{gh}^{[2]}\[v, \Xi; \Omega\] + \^R_k\)_{\!\xi   \bar\xi' }^{-1}
          \(\partial_t \^R_k\)_{\!\xi   \bar\xi' }\]                        \\
 &-& \Tr\[\(S_\mr{gh}^{[2]}\[v, \Xi; \Omega\] + \^R_k\)_{\!\xi'  \bar\xi  }^{-1}
          \(\partial_t \^R_k\)_{\!\xi'  \bar\xi  }\]                        \\
 &-& \Tr\[\(S_\mr{gh}^{[2]}\[v, \Xi; \Omega\] + \^R_k\)_{\!\beta \bar\beta}^{-1}
          \(\partial_t \^R_k\)_{\!\beta \bar\beta}\]
                \end{eqnarray*}}\hfill
\parbox{7mm}{\equa{4.6}{}}
\newline
where $\gam_k\!\[v; \Omega\] \equiv \bar\gam_k + S_\mr{gf}$. The derivatives 
have to be taken at the point $v = \Xi = 0$. 

We see that the l.\,h.\,s.~of the truncated evolution equation does not 
contain any  deviations from the ghosts since their evolution has been 
neglected. But of course, the r.\,h.\,s.~contains such contributions, namely 
the last three traces. Here the approximation leads to the appearance of the 
\emph{free} regularized propagators $(S_\mr{gh}^{[2]} + \^R_k)^{-1}$ instead 
of the full ones.


\ssection{Evolution Equation for the N=1 SYM Coupling Constant}
As a simple application let us specify an explicit ansatz for 
$\gam_k\!\[v; \Omega\]$ and calculate the $\beta$-function of pure 
N=1 SYM theory. We use the simplest nontrivial possibility:
\newline
\parbox{13.9cm}{\begin{eqnarray*}
         \gam_k\[v; \Omega\] =
     &-& \frac{Z_k}{2 g^2}\,
            \Tr\!\[ \(e^{-v} \nab^\beta e^v\) \cdot \bar \nab^2\!\!
            \(e^{-v} \nab_{\!\beta} e^v\) \]                           \\
        &-& \frac{Z_k}{2 \alpha g^2}\,
            \Tr\!\[ v \(\nab^2 \bar \nab^2 + \bar \nab^2 \nab^2 \)v \].
                \end{eqnarray*}}\hfill
\parbox{7mm}{\equa{5.1}{}}
\newline
The first term is the classical invariant action where we have replaced the 
bare coupling constant $g^2$ by the renormalized one 
$g_k^2 \equiv Z_k^{-1} g^2$. We will not consider here the other 
renormalization constants. In principle, this is possible but it would go 
beyond the aim of this paper. The second term is the gauge fixing.

This leads to the following inverse propagator for the gauge field
\newline
\parbox{13.9cm}{\begin{eqnarray*}
                \(\gam_k^{[2]}\)_{\! v v}
      &=& - \frac{Z_k}{g^2}\[-\nab^\alpha \bar\nab^2 \nab_\alpha
          - W^\alpha \nab_\alpha + \frac{i}{2}\( \nab^\alpha W_\alpha\)
          + \frac{1}{\alpha}\(\nab^2 \bar\nab^2 + \bar\nab^2 \nab^2\)\]  \\
 &\equiv& - \frac{Z_k}{g^2}\[ \BBox_V + \(\frac{1}{\alpha} - 1\)
            \(\nab^2 \bar\nab^2 + \bar\nab^2 \nab^2\)\]
                \end{eqnarray*}}\hfill
\parbox{7mm}{\equa{5.2}{}}
\newline
Here $W^\alpha$ is the background field strength. To simplify further we 
restrict ourselves to the case where the gauge parameter has been set 
$\alpha = 1$. If $\alpha$ would depend on the scale $k$ it is shown in 
\cite{t} --- for a non-covariant gauge in non-SUSY gauge theory --- that 
$\alpha=0$ is a fixed point of the evolution. The arguments used there can 
be extended to general linear covariant gauges and to SYM. Physical 
quantities are independent from the choice of the gauge parameter.

Calculating the contributions from the ghost terms we find
\equ{5.3}{\(S_\mr{gh}^{[2]}\)_{\!\xi   \bar\xi' } =
          \(S_\mr{gh}^{[2]}\)_{\!\xi'  \bar\xi  } =
          \(S_\mr{gh}^{[2]}\)_{\!\beta \bar\beta} = - \nab^2 \bar\nab^2 .}
With this the evolution equation (\ref{e4.6}) can be written as
\newline
\parbox{13.9cm}{\begin{eqnarray*}
          \partial_t \gam_k
 &=& \;\, \Tr \!\[\Big( - \BBox_V           + k^2 R^{(0)}\Big)^{-1}\:
          \frac{\partial_t\(   k^2 Z_k R^{(0)}   \)}{Z_k} \]       \\
 &-& 3    \Tr \!\[\Big( - \nab^2 \bar\nab^2 + k^2 R^{(0)}\Big)^{-1}\:
                \partial_t\Big(k^2     R^{(0)}\Big)       \] .
                \end{eqnarray*}}\hfill
\parbox{7mm}{\equa{5.4}{}}
\newline
Because we did not gauge the symmetry (\ref{e2.8.1}) we restricted the 
trace in the last term to the antichiral subspace where $\nab^2 \bar\nab^2$ 
is invertible. This fact will become obvious when we calculate the traces.

In addition we have chosen the cutoff operators as
\equ{5.5}{R_k^\mr{YM} = g^{-2} k^2 Z_k\, R^{(0)} \!\( - \frac{\BBox_V}{k^2} \),
\quad     R_k         =        k^2       R^{(0)} \!\( - \frac{\nab^2
          \bar\nab^2}{k^2} \)}
with $R^{(0)}$ as in (\ref{e2.9.1}) such that the modes are cut off in a 
proper way as required. Because of our simple ansatz the propagators in 
(\ref{e5.4}) have the form of free (regularized) propagators. This means 
that this truncated evolution equation is very near to an one-loop equation.

In (\ref{e5.4}) we have taken all quantities at $e^{\bar\Omega} e^\Omega =
e^{v_\mr{S}}$. This shows the freedom in choosing a specific splitting. Our
aim is to make an expansion of (\ref{e5.4}) with respect to
$\chint W^\alpha W_\alpha$ which must have a value being non-zero. On the 
other hand, the calculation should be simplified as much as possible. 
Therefore it is very convenient to choose 
$e^{\bar\Omega} e^\Omega = e^{v_\mr{S}}$.

For the l.\,h.\,s.~of (\ref{e5.4}) we find at 
$e^{\bar\Omega} e^\Omega = e^{v_\mr{S}}$
\equ{5.6}{\mr{l.\,h.\,s.} = \frac{\partial_t Z_k}{2 g^2} \: 
          \chint W^\alpha W_\alpha}
where $W^\alpha$ is the background field strength. On the r.\,h.\,s.~of 
(\ref{e5.4}) we use the heat kernel expansion to calculate the traces 
\cite{m}. 

Let us define the heat kernel coefficients $a_n$ with
\equ{5.7}{\iint \sum_{n=0}^{\infty} a_n (z) \(i s\)^\frac{n -4}{2} = i K (s)}
where the kernel is defined as
\equ{5.8}{K(s) = \vint\lim_{z \to z'} e^{- i s\, O(z)} \:\delta^{(8)} (z - z').}
The operator $O$ can be chosen as $\BBox_V$ or $\nab^2 \bar\nab^2$.

Let us consider the case $O = \nab^2 \bar\nab^2$. Using the total chiral 
representation we extract from the volume element the derivatives 
$\vint = \chint \bar\nab^2$. Then we find the chiral kernel
\equ{5.9}{K(s) = \chint \lim_{z \to z'} e^{- i s\, \bar\nab^2 \nab^2} \:
         \delta^{(6)} (z - z')}
with the chiral $\delta$-distribution $\delta^{(6)}(z) = \bar\nab^2
\delta^{(8)}(z)$. Because $K(s)$ is chiral in (\ref{e5.9}) we can replace
$\bar\nab^2 \nab^2$ by the background covariant chiral Laplacian
$$\BBox_+ \equiv \bar\nab^2 \nab^2 + \nab^2 \bar\nab^2 - \bar\nab^{\dot\alpha}
  \nab^2 \bar\nab_{\dot\alpha} = \BBox  - i\, W^\alpha \nab_\alpha -
  \frac{i}{2} \( \nab^\alpha W_\alpha\)$$
where $\BBox \equiv \nab^{\alpha \dot\alpha} \nab_{\alpha \dot\alpha}$.

Now we want to perform the coincidence limit. To this purpose we need the 
momentum representation of $\delta$:
\equ{5.10}{\delta^{(6)} (z-z') = \int \!\! \frac{d^{\;\! 4} k}{\(2 \pi\)^4} \:
           e^{i k_a \(x_a - x_a' - \frac{i}{2}\(\theta'^\alpha
           \bar\theta^{\dot\alpha} + \bar\theta'^{\dot\alpha} \theta^\alpha\)\)}
        \int\!\! d^{\;\! 2} \eps\: e^{i \eps^\alpha \(\theta_\alpha -
        \theta'_\alpha\)} .}
The supermomentum $\eps$ is the Fourier transform of the supercoordinate 
$\theta$. Equation (\ref{e5.10}) corresponds to the translation  
$(x, \theta, \bar\theta) \to (x', \theta', \bar\theta')$ in the superspace. 
With this the chiral kernel reads after a rescaling
\newline
\parbox{13.9cm}{\begin{eqnarray*}
               K(s)
       &=&    \frac{1}{s^2}\, \chint\!\!\! \int\!\! 
              \frac{d^{\;\! 4} k}{\(2 \pi\)^4}\,
              d^{\;\! 2} \eps \: e^{i k^2} \:
              \exp \! \[- i\,s\, \BBox + 2 s^{1/2} k^a \nab_a -
              \frac{1}{2} s \(\nab^\alpha W_\alpha\) \right.             \\
       & &    \hspace{10.3em} \left.   - s\, W^\alpha \nab_\alpha +
              \frac{1}{2} s^{1/2} W^\alpha k_\alpha{}^{\dot\alpha}\,
              \bar\theta_{\dot\alpha} - i\, s\, W^\alpha \eps_\alpha\] .
                \end{eqnarray*}}\hfill
\parbox{7mm}{\equa{5.11}{}}
\newline
We are searching for the first nonvanishing coefficient. Because $\eps$ is a
Grassmann variable the integration over $d^{\;\! 2} \eps$ requires a term
proportional to $\eps^2$ to be nonvanishing. If we expand the second 
exponential function with respect to $s$ it becomes clear that this term is 
at least of order $s^2$. The complete contribution to $K(s)$ from this term 
is proportional to $s^0$ or higher. From (\ref{e5.7}) it follows that $a_4$ is 
proportional to $s^0$ which is therefore the lowest nonvanishing coefficient:
$$\iint a_4 (z) = \frac{i}{2}\, \frac{1}{16 \pi^2}\, \chint W^\alpha W_\alpha.$$
Therefore the trace of the chiral kernel in lowest order is
\equ{5.12}{\Tr \, e^{-i\,s \nab^2 \bar\nab^2} = \frac{1}{2}\, 
           \frac{1}{16 \pi^2}\, \chint W^\alpha W_\alpha + O(s^{1/2}) .}

An analogous calculation for $\BBox_V$ leads to the result that the deviation 
from the trace of its kernel can be neglected because it is of higher order 
in $s$:
\equ{5.13}{\Tr \, e^{-i\, s\, \BBox_V} = O(s^2) .}
The first nonvanishing coefficient is $a_8$ which is proportional to
$W^2 \, \bar W^2$. In the framework of our truncation this is negligible.

For a general function $F(-\nab^2 \bar\nab^2)$ we obtain for the trace in the
framework of our approximation
\equ{5.14}{\Tr \!\[F (-\nab^2 \bar\nab^2)\] = \frac{N}{2}\, \frac{1}{16 \pi^2}
           F(0)\, \chint W^\alpha W_\alpha .}
Here we have taken into account the space of fields where $\nab^2 \bar\nab^2$
acts. The ghost and gauge fields are tensors of second order in the 
representation of the gauge group. This leads to a factor of the dimension of 
this representation, which is equal to the dimension of $SU(N)$ in the 
adjoint representation used here.

With this result we go back to the truncated evolution equation (\ref{e5.4}) 
to express the r.\,h.\,s.~as
\equ{5.15}{\mr{r.\,h.\,s.} = -3 \,\frac{N}{16 \pi^2} \, \chint W^\alpha 
           W_\alpha.}
Comparing (\ref{e5.6}) and (\ref{e5.15}) and dividing them by $Z_k$ we find 
the anomalous dimension of $\chint W^\alpha W_\alpha$
\equ{5.16}{\eta_k \equiv - \frac{\partial_t Z_k}{Z_k} = N\,\frac{3}{8 \pi^2}\,
           g_k^2 .}
Together with $\partial_t\, g_k^2 = \partial_t\, Z_k^{-1} \, g^2 \equiv 
\beta (g_k)$ we find the evolution equation for the N=1 SYM coupling constant
\equ{5.17}{\partial_t\, g_k^2 = \eta_k\, g_k^2 =  N\, \frac{3}{8 \pi^2}\,
           g_k^4}
which we intended to derive. Their solution
\equ{5.18}{g_k^2 = g_0^2 \,\frac{1}{1 - \frac{3}{8 \pi^2}\, N \, g_0^2 \ln\!
           \(k / k_0\)} }
fixes the action (\ref{e5.1}) uniquely.

The r.\,h.\,s.~of (\ref{e5.17}) is the well-known one-loop $\beta$-function 
of pure N=1 SYM theory \cite{o}. It is remarkable that neither it depends 
on the specific shape of the cutoff function $R^{(0)}$ nor on the gauge 
parameter. This expresses the scheme independence of the one-loop contribution 
to the $\beta$-function which is in accordance with refs.~\cite{o} where very 
different methods, such as loop calculations with dimensional regularization or 
the instanton calculus, have been used.

Normally, even a simple ansatz such as (\ref{e5.1}) leads not only to 
one-loop contributions but also to higher ones (see e.\,g.~\cite{e}). In SYM 
theory, because of (\ref{e5.13}), this is not the case. This is an indication 
of how non-renormalization theorems could work within this approach.

But to demonstrate the full power of our method it would be desirable to do 
the calculation with an improved ansatz allowing for the renormalization of 
additional fields and containing higher invariants such as $W^2 \bar W^2$ 
which are not out of the question to be generated during the evolution. 
Such an analysis will bring a lot of technical problems, e.\,g.~it is not 
clear how to calculate the traces of the appearing operators. In addition, 
considering e.\,g.~finite theories, one can study non-renormalization 
theorems.


\ssection{Conclusion}
We generalized the effective average action to the superspace formalism which
opens the possibility to do any calculation in a manifest supersymmetric way.
Using as an example the N=1 SYM theory, we defined the effective average 
action (\ref{e2.12}). With the modified STI (\ref{e3.4}) we found a 
possibility to control the breaking of the BRST invariance of $\gam_k$. Then 
we derived the exact evolution equation (\ref{e4.1}). Because we were not 
able to solve this equation exactly, we used a truncation (\ref{e4.2}) to 
simplify (\ref{e4.1}) leading to the truncated evolution equation 
(\ref{e4.6}). As an example of a possible truncation we made the ansatz 
(\ref{e5.1}) to calculate the one-loop contribution to the $\beta$-function 
in N=1 SYM theory (\ref{e5.16}).

To extend the aim of the paper it would be interesting to generalize $\gam_k$ 
to theories with multiple SUSY. In principle, this should cause no great 
difficulties.
In addition, it would be very interesting to consider supergravities. Then 
the effective average action has to be applied to local supersymmetries. This 
may be not unrealistic since in \cite{k, l} the effective average action has 
been considered in (Einstein) gravity, partially with additional fields (all 
spins from $s=0$ to $s=2$). On the other hand, we have shown here that a 
cutoff can be introduced without breaking (global) SUSY. There is no 
contradiction between these works preventing a consistent cutoff in 
supergravity.

\emph{Note added}: After the publishing of this article in hep-th Bonini 
et.\,al.~published a paper \cite{u} about Wilson renormalization group in 
SUSY. Their approach is in the spirit of \cite{i} emphasising the 
perturbative features of Wilson RG.

\section*{Acknowledgments}
We thank M.~Reuter as well as U.~Ellwanger and J.~Pawlowski for helpful and 
enlightening discussions. S.\,F.~has been supported by DFG, project GRK 52.


\end{document}